\def\fr#1#2{\hbox{${#1\over #2}$}}
\def\prg#1{\medskip{\bf #1}}
\def\lra{\leftrightarrow}           
\def\ni{\noindent}                  \def\pd{\partial}

\def\eps{{\epsilon}}

       \def\bL{{\bar L}}

     \def\tG{{\tilde G}}
       \def\tom{\tilde\omega}

\def\m{\mu}             \def\n{\nu}              
\def\G{\Gamma}          \def\g{\gamma}           \def\d{\delta}
          \def\s{\sigma}           \def\t{\tau}
\def\a{\alpha}          \def\b{\beta}            \def\th{\theta}
\def\vphi{\varphi}      \def\ve{\varepsilon}     \def\p{\pi}
\def\r{\rho}            \def\D{\Delta}           \def\L{\Lambda}
\def\l{\lambda}         \def\om{\omega}          \def\Om{\Omega}

\def\cL{{\cal L}}       \def\cH{{\cal H}}        
\def\cM{{\cal M }}      \def\cO{{\cal O}}        \def\cE{{\cal E}}
\def\cK{{\cal K}}
\def\hcO{{\hat\cO}}     \def\hcH{{\hat\cH}}
\def\tH{\tilde H}       \def\bcH{{\bar\cH}}
\def\bcK{{\bar\cK}}     \def\bD{{\bar D}}

\def\Leff{\hbox{$\L_{\rm eff}$}}
\def\dis{\displaystyle}
\def\nn{\nonumber}
\def\be{\begin{equation}}             \def\ee{\end{equation}}
\def\ba#1{\begin{array}{#1}}           \def\ea{\end{array}}
\def\bea{\begin{eqnarray} }           \def\eea{\end{eqnarray} }
\def\beann{\begin{eqnarray*} }        \def\eeann{\end{eqnarray*} }
\def\lab#1{\label{eq:#1}}             \def\eq#1{(\ref{eq:#1})}
\def\bsubeq{\begin{mathletters}}      \def\esubeq{\end{mathletters}}
\def\bitem{\begin{itemize}}           \def\eitem{\end{itemize}}
\renewcommand{\arraystretch}{1.1}

\documentstyle[aps,preprint,eqsecnum]{revtex}        

\begin{document}
\tighten

\title{Asymptotic symmetries in 3d gravity with torsion}

\author{M.\ Blagojevi\'c$^{1,2}$ and M. Vasili\'c$^{1,}$\thanks{Email
        addresses:  mb@phy.bg.ac.yu, mvasilic@phy.bg.ac.yu}}

\address{${}^1$Institute of Physics, P.O.Box 57, 11001 Belgrade,
             Yugoslavia\\
         ${}^2$Primorska Inst. for Natural Sci. and Technology,
         P.O.Box 327, 6000 Koper, Slovenia}


\maketitle

\begin{abstract}
We study the nature of asymptotic symmetries in topological 3d gravity
with torsion.  After introducing the concept of asymptotically anti-de
Sitter configuration, we find that the canonical realization of the
asymptotic symmetry is characterized by the Virasoro algebra with
classical central charge, the value of which is the same as in general
relativity: $c=3\ell/2G$.

\end{abstract}

\section{Introduction}

Three-dimensional (3d) gravity has been used as a theoretical
laboratory to test some of the conceptual problems of quantum gravity.
The phase space structure of 3d gravity is known to be of great
importance not only at the classical level, but also for a clear
understanding of the related quantum structure \cite{1,2}. In
particular, the Virasoro algebra of the asymptotic symmetry plays a
central role in our understanding of the quantum nature of black hole
\cite{3,4,5,6,7,8,9,10}. One can observe, however, that the analysis of
these issues has been carried out only in {\it Riemannian\/} spacetime
of general relativity. In the present paper we begin an investigation
of the asymptotic structure of 3d gravity in the context of {\it
Riemann--Cartan\/} geometry, a geometry possessing a metric compatible
connection, with both the curvature and the torsion of the underlying
spacetime manifold \cite{11,12}. In this way, we expect to clarify the
influence of spacetime geometry on the boundary dynamics.

Dynamics of a theory is determined not only by the action, but also by
the asymptotic conditions. The dynamical role of asymptotic conditions
is best seen in topological theories, where the non-trivial dynamics is
bound to exist only at the boundary. General action for topological 3d
gravity in Riemann--Cartan spacetime has been proposed by Baekler and
Mielke \cite{13,14}. For particular values of parameters, this action
leads to the {\it teleparallel\/} (Weizenb\"ock) geometry, defined by
the requirement of vanishing curvature \cite{15,16,17,12}. Teleparallel
geometry is, in a sense, complementary to Riemannian: curvature
vanishes, and torsion remains to characterize the parallel transport.
We choose this teleparallel framework to study the asymptotic structure
of spacetime in the presence of torsion. After introducing the concept
of asymptotically anti-de Sitter (AdS) field configuration and
performing the canonical analysis, we find that the teleparallel
spacetime has the same asymptotic structure as the Riemannian spacetime
in general relativity.

The paper is  organized as follows. In Sect. II we introduce
Riemann--Cartan spacetime as a general geometric arena for 3d gravity
with torsion, and show that there is a specific choice of parameters
which leads to the teleparallel description of gravity. In Sect. III we
construct two exact solution of the resulting teleparallel theory: the
AdS solution and the black hole. Then, in Sect. IV, we introduce the
concept of asymptotically AdS configuration. The symmetry of such a
configuration, the asymptotic symmetry, is shown to be the same as in
general relativity --- the conformal symmetry. In the next two sections,
the gauge structure of the theory is incorporated into the canonical
formalism. First, the general Hamiltonian structure of the theory is
derived in Sect. V. After that, in Sect. VI, we construct the canonical
gauge generators compatible with the adopted asymptotic conditions. The
construction is realized with the help of appropriate boundary terms
\cite{18}, which are interpreted as the conserved charges of the
theory. The investigation of the Poisson bracket algebra of the
asymptotic generators leads to the central result of the paper: the
asymptotic symmetry is characterized by the classical Virasoro algebra
with central charge, the value of which is the same as in Riemannian
spacetime of general relativity: $c=3\ell/2G$ \cite{4,5,6,7,8,9,10}.
Finally, section VII is devoted to concluding remarks, while Appendices
contain some technical details.

Our conventions are given by the following rules: the Latin indices
refer to the local Lorentz frame, the Greek indices refer to the
coordinate frame; the first letters of both alphabets
$(a,b,c,...;\a,\b,\g,...)$ run over 1,2, the middle alphabet letters
$(i,j,k,...;\m,\n,\l,...)$ run over 0,1,2; the tetrad field $b^i{_\m}$
and its dual $h_i{^\m}$ are used to convert Greek and Latin indices
into each other; $\eta_{ij}=(+,-,-)$ and
$g_{\m\n}=b^i{_\m}b^j{_\n}\eta_{ij}$ are the metric components in the
tangent and coordinate frame; totally antisymmetric tensor $\ve^{ijk}$
and the related tensor density $\ve^{\m\n\r}$ are both normalized so
that $\ve^{012}=1$.

\section{Basic dynamical features}

\prg{1.} Three-dimensional gravity with torsion can be formulated as
Poincar\'e gauge theory, with an underlying  geometric structure
described by {\it Riemann--Cartan\/} space $U_3$. Basic gravitational
variables of the theory are the triad field $b^i{_\m}$ and the Lorentz
connection $A^{ij}{_\m}$. Their gauge transformations are local Lorentz
rotations and local translations, with the parameters $\ve^{ij}$ and
$\xi^\m$, respectively \cite{11,12}:
\bea
&&\d_0 b^i{_\m}=\ve^i{_k}b^k{_\m}-\xi^\r{}_{,\m}b^i{_\r}
                               -\xi^\r\pd_\r b^i{_\m}\, ,    \nn\\
&&\d_0 A^{ij}{_\m}=-\nabla_\m\ve^{ij}-\xi^\r{}_{,\m}A^{ij}{_\r}
                               -\xi^\r\pd_\r A^{ij}{_\m}\, . \nn
\eea
The related field strengths $T^i{}_{\m\n}$ and $R^{ij}{}_{\m\n}$ are
geometrically identified with the torsion and the curvature:
\bea
&&T^i{}_{\m\n}=\pd_\m b^i{_\n}+ A^i{}_{m\m}b^m{_\n}-(\m\lra\n)\,,\nn\\
&&R^{ij}{}_{\m\n}=\pd_\m A^{ij}{_\n}+A^i{}_{m\m}A^{mj}{_\n}
                                                   -(\m\lra\n)\, .\nn
\eea
Note that, in this approach, metric is not an independent field variable:
$g_{\m\n}$ is defined in terms of $b^i{_\m}$ and the tangent space
metric $\eta$ by the relation $g_{\m\n}=b^i{_\m}b^i{_\n}\eta_{ij}$.

In $d=3$, we can simplify the notation by introducing
\bea
&&\om_{i\m}=-\fr{1}{2}\ve_{ijk}A^{jk}{_\m}\, ,\qquad
  \th_i=-\fr{1}{2}\ve_{ijk}\ve^{jk}  \, ,                 \nn\\
&& R_{i\m\n}=-\fr{1}{2}\ve_{ijk}R^{jk}{}_{\m\n}\, .       \nn
\eea
Then, the transformation laws of the gauge fields take the form
\bsubeq
\bea
\d_0 b^i{_\m}&=& \ve^{ijk}\th_j b_{k\m}-\xi^{\r}{}_{,\,\m}b^i{_\r}
               -\xi^{\r}\pd_\r b^i{}_{\m}                   \nn\\
\d_0\om^i{_\m}&=& -\nabla_\m\th^i-\xi^{\r}{}_{,\,\m}\om^i{_\r}
                  -\xi^{\r}\pd_\r\om^i{}_{\m}\, ,        \lab{2.1a}
\eea
where $\nabla_\m\th^i=\pd_\m\th^i+\ve_{imn}\om^m{_\m}\th^n$, and the
field strengths are given as
\bea
&&T^i{}_{\m\n}=\pd_\m b^i{_\n}
               +\ve^{ijk}\om_{j\m}b_{k\n}-(\m\lra\n)\, , \nn\\
&&R_{i\m\n}=\pd_\m\om_{i\n}-\pd_\n\om_{i\m}
               +\ve_{ijk}\om^j{_\m}\om^k{_\n}\, .        \lab{2.1b}
\eea
\esubeq

\prg{2.} Dynamical structure of the theory is determined by an action
integral (the important role of boundary conditions will be discussed
later). Direct generalization of Einstein's theory to the $U_3$ space
leads to Einstein--Cartan theory:
$$
I_{EC}=-a\int d^3x bR\, ,\qquad a=\frac{1}{16\pi G}\, .
$$
Staying in the realm of Riemannian geometry (vanishing torsion),
Witten \cite{2} demonstrated that Einstein--Cartan theory with
cosmological constant is equivalent to the standard gauge theory of
Chern--Simons type. An interesting extension of these ideas to
Riemann--Cartan space (non-vanishing torsion) has been proposed by
Baekler and Mielke \cite{13,14}. They studied an action constructed out
of the following topological or topological-like terms:
\bsubeq\lab{2.2}
\bea
I_1&=&-\int d^3x b(aR+2\L)
     =\int d^3x\ve^{\m\n\r}\left(ab^i{_\m}R_{i\n\r}
  -\fr{1}{3}\L\ve_{ijk}b^i{_\m}b^j{_\n}b^k{_\r}\right)\, ,    \nn\\
I_2&=&\int d^3x\ve^{\m\n\r}\left( \om^i{_\m}\pd_\n\om_{i\r}
 +\fr{1}{3}\ve_{imn}\om^i{_\m}\om^m{_\n}\om^n{_\r} \right)\, ,\nn\\
I_3&=&\fr{1}{2}\int d^3x\ve^{\m\n\r}b^i{_\m}T_{i\n\r}\, ,   \lab{2.2a}
\eea
where $\L$ is a constant. The first term describes
Einstein--Cartan theory with cosmological constant, the second term is
the Chern--Simons action for the Lorentz connection, and the third term
represents an action of the translational Chern--Simons type \cite{19}.
The general Baekler--Mielke action reads:
\be
I=I_1+\a_2 I_2+\a_3 I_3 +I_M\, ,                            \lab{2.2b}
\ee
\esubeq
where $I_M$ is an action for matter fields.

Varying the action with respect to $b^i{_\m}$ and $\om^i{_\m}$, we
obtain the field equations:
\bea
&&\ve^{\m\n\r}\left[ aR_{i\n\r}-\L\ve_{ijk}b^j{_\n}b^k{_\r}
   +\a_3 T_{i\n\r}\right]=\t^\m{_i}\, ,   \nn\\
&&\ve^{\m\n\r}\left[ aT_{i\n\r}+\a_2 R_{i\n\r}
   +\a_3\ve_{ijk}b^j{_\n}b^k{_\r} \right]=\s^\m{_i}\, ,     \lab{2.3}
\eea
where $\t$ and $\s$ are the energy-momentum and spin tensors of
matter fields, respectively. From these equations one can calculate
explicitly the torsion and the curvature of the spacetime. For
our purposes, it is sufficient to consider only the solutions of these
equations in vacuum, where $\t=\s=0$. In that case, and for
$\a_2\a_3-a^2\ne 0$, we find
\be
T_{ijk}=A\ve_{ijk}\, ,\qquad R_{ijk}=B\ve_{ijk}\, ,         \lab{2.4}
\ee
with
$$
A\equiv \frac{\a_2\L+\a_3 a}{\a_2\a_3-a^2}\, ,\qquad
B\equiv -\frac{(\a_3)^2+a\L}{\a_2\a_3-a^2}\, .
$$

In Riemann--Cartan space $U_3$, the Lorentz connection can be expressed
in terms of the Levi--Civita connection $\D$ and the contortion $K$ as
$A=\D+K$ \cite{11,12}. Substituting this expression into the definition
of the curvature tensor $R^{ij}{}_{\m\n}(A)$ leads to the geometric
identity
$$ R^{ij}{}_{\m\n}(A)=R^{ij}{}_{\m\n}(\D)+
  \left[\nabla_\m K^{ij}{_\n}-K^i{}_{s\m}K^{sj}{_\n}
                             -(\m\lra\n)\right]\, .
$$
Then, by combining this identity with the vacuum field equations
\eq{2.4}, we obtain the following expression for the Riemannian piece of
the $U_3$ curvature:
\be
R^{ij}{}_{\m\n}(\D)=-\Leff(b^i{_\m}b^j{_\n}-b^i{_\n}b^j{_\m})\, ,
\qquad \Leff\equiv B-\frac{1}{4}A^2 \, ,                   \lab{2.5}
\ee
where $\Leff$ is the effective cosmological constant. Looking at this
equation as an equation for the metric, we see that the metric of our
spacetime is maximally symmetric; for $\Leff<0$ ($\Leff>0$) it has the
anti-de Sitter (de Sitter) form.

\prg{3.} At the end of this section, we would like to comment on two
special cases of the Baekler--Mielke action.

For $\a_2=\a_3=0$ (Witten's choice \cite{2}), we have
\be
T_{ijk}=0\, , \qquad
R_{ijk}=\frac{\L}{a}\,\ve_{ijk}\, .                        \lab{2.6}
\ee
The torsion vanishes, and the geometry of spacetime becomes
{\it Riemannian\/}.

Another interesting choice is $(\a_3)^2+a\L=0$. It yields the field
equations
\setcounter{equation}{7}
$$
T_{ijk}=-\frac{\a_3}{a}\,\ve_{ijk}\, ,  \qquad
R_{ijk}=0 \, ,                             \eqno\hbox{\rm (2.7a,b)}
$$
which are ``geometrically dual" to those of Witten: the curvature
vanishes, and the geometry becomes {\it teleparallel\/}.

Having in mind our intention to study the role of torsion in the
boundary dynamics, we restrict our attention to the teleparallel case
(2.7). Since the field equations are independent of $\a_2$, we also
assume $\a_2=0$. The effective cosmological constant is now negative:
\bsubeq
\be
\Leff=-\frac{1}{4}A^2\equiv-\frac{1}{\ell^2}<0\, .         \lab{2.8a}
\ee
After introducing the constant $\ell$ by the relation $A=2/\ell$,
these conditions are summarized as
\be
\a_2=0\, ,\qquad \a_3=-\frac{2a}{\ell}\, ,
          \qquad \L=-\frac{4a}{\ell^2}\, ,                 \lab{2.8b}
\ee
\esubeq
and the general action \eq{2.2b} in the absence of matter reduces to
the form
\be
I= -a\int d^3x\, b\left( R+\frac{1}{\ell}\,\ve^{ijk}\,T_{ijk}
                       -\frac{8}{\ell^2}\right)\, .        \lab{2.9}
\ee

\section{Exact vacuum solutions}

We now direct our attention to the exact classical solutions of the
vacuum field equations (2.7). In this regard, it is useful to note that
equations (2.7b) and \eq{2.5} are equivalent, provided equation (2.7a)
holds. As a consequence, our search for the exact solutions will be
based on the following strategy:
\bitem
\item[] i) we shall first find a solution of equation \eq{2.5} for the
metric; \\
ii) given the metric, we shall procede to find a solution for
the triad field;
\\ iii) finally, we shall use equation (2.7a) to determine the
connection.
\eitem
After that, equation (2.7b) will be automatically satisfied.

The first step in the above procedure is very simple, since the form of
the metric in maximally symmetric 3d spaces is well known \cite{20}.

\subsection{Teleparallel AdS solution}

As the first solution of \eq{2.5} with $\Leff=-1/\ell^2$, we display
the metric of the AdS solution in static coordinates $x^\m=(t,r,\vphi)$:
\be
ds^2=f^2dt^2 -f^{-2}dr^2-r^2d\vphi^2\, ,\qquad
f^2\equiv 1+\frac{r^2}{\ell^2}\, .                        \lab{3.1}
\ee
The related triad field can be chosen to have the simple, diagonal form:
\be
b^0=fdt\, , \qquad b^1=f^{-1}dr \, ,
              \qquad  b^2=rd\vphi\, ,                     \lab{3.2}
\ee
where $b^i=b^i{_\m}dx^\m$. It produces the metric \eq{3.1} via
$ds^2=b^i b^j\eta_{ij}$.

The connection is determined from equation (2.7a). Introducing the
light-cone coordinates
$$
x^\pm=\frac{1}{\ell}x^0\pm x^2\, ,
$$
the solution for the connection 1-form $\om^i=\om^i{_\m}dx^\m$ is
given as
\be
\om^0=fdx^-\, ,\qquad
\om^1=\frac{1}{\ell f}\,dr\, , \qquad
\om^2=-\frac{r}{\ell}\,dx^- \, .                          \lab{3.3}
\ee

Thus, equations \eq{3.2} and \eq{3.3} represent the exact AdS vacuum
solution of our theory. This solution is defined in the realm of the
teleparallel geometry, and should not be confused with the AdS solution
\eq{3.1} in Riemannian geometry.

The symmetries of the AdS solution are discussed in Appendix A.

\prg{Comment.} Given the metric \eq{3.1}, the choice of the AdS pair
$(b^i{_\m},\om^i{_\m})$ is not unique, in the sense that any Lorentz
transform of a particular solution yields also a solution of the
theory. On the other hand, as a consequence of $R_{i\m\n}=0$, there
exists a solution with the trivial connection $\tom^i{_\m}=0$. Any other
vacuum connection can be written as a Lorentz transform of
$\tom^i{_\m}$. This is especially true for our vacuum connection
\eq{3.3}:
$$
-\ve^{ijk}\om_{k\m}=\L^i{_k}\,\L^{jk}{}_{,\,\m}\ ,\qquad
\L^i{_k}\L^j{_l}\,\eta^{kl}=\eta^{ij} \, .
$$
From this equation, we find the following particular solution for the
Lorentz matrix $\L$:
$$
\L^i{_j}=
  \left( \ba{ccc}
  -f & \dis\frac{r}{\ell}\sin x^- &
  \dis-\frac{r}{\ell}\cos x^-                   \\
   0 & -\cos x^-                   & -\sin x^-  \\
  \dis\frac{r}{\ell} & -f\sin x^-  & f\cos x^-
         \ea  \right) \, .
$$
The general solution is obtained by the replacement
$\L\to\L\L_{\rm c}$, where $\L_{\rm c}$ is a constant Lorentz matrix.

Now, when we know the Lorentz matrix which transforms the trivial
connection $\tilde\om^i_\m=0$ into our $\om^i{_\m}$, we can easily
find the related triad field
$\tilde b^i{_\m}$ as $\,\tilde b^i{_\m}=\L_k{^i}\,b^k{_\m}\,$:
{\renewcommand{\arraystretch}{1.6}
$$
\tilde b^i{_\m}=\left(
\ba{ccc}
-f^2 & 0 & \dis-\frac{r^2}{\ell}  \\
\dis-\frac{r}{\ell}\,f\sin x^- & \dis-f^{-1}\cos x^- & -rf\sin x^-  \\
\dis\frac{r}{\ell}\,f\cos x^- & \dis-f^{-1}\sin x^- & rf\cos x^-
\ea
\right) \,.
$$ }

\ni Thus, $\tilde b^i{_\m}$ and $\tilde\om^i{_\m}=0$ are also vacuum
solutions of the field equations (2.7), which differ from those given
in \eq{3.2} and \eq{3.3} by the local Lorentz rotation.

Our vacuum triad \eq{3.2} is not well defined at $r=0\,$, while
$\,\tilde b^i{_\m}\,$ is. If we are only interested in the asymptotic
region (as we are), both vacuum solutions are acceptable.

\subsection{Teleparallel black hole}

Another well known solution of equation \eq{2.5} is the BTZ black
hole \cite{21}. In static coordinates $(t,r,\vphi)$, the black hole
is defined by the metric (in units $4G=1$)
\bea
&&ds^2=N^2dt^2-N^{-2}dr^2-r^2(d\vphi+N_\vphi dt)^2\, ,\nn\\
&&N^2\equiv\left(-2m+\frac{r^2}{\ell^2}+\frac{J^2}{r^2}\right)\, ,
\qquad N_\vphi\equiv\frac{J}{r^2}\, ,                        \lab{3.4}
\eea
with $0\le\vphi<2\pi$.
Although the AdS vacuum and the black hole are locally isometric
solutions, they are globally distinct: the black hole describes a
conic geometry obtained by a geometric identification of points in
AdS space \cite{21}. The two parameters $m$ and $J$ are related to
the global properties of conic geometries known as "missing angle"
and "time jump" \cite{1}. The thorough analysis of Ref. \cite{21}
shows that all physically acceptable solutions of our theory are
exhausted by the two-parameter black hole solution \eq{3.4}. As we
shall see later, the parameters $m$ and $J$ have the physical
meaning of energy and angular momentum.

The black hole triad and connection are not uniquely defined by the
metric \eq{3.4}. Although all possible solutions are locally
equivalent, they may differ globally. It can be shown, for example,
that the solution with everywhere vanishing connection {\it is not}
globally well defined for all the values of $m$ and $J$.

In what follows, we adopt the simple ansatz for the triad field
(see also Ref. \cite{22}):
\bea
&&b^0=Ndt \, ,\qquad b^1=N^{-1}dr \, ,                   \nn\\
&&b^2=r(d\vphi+N_\vphi dt)\, .                          \lab{3.5}
\eea
The connection is, again, determined by equation (2.7a),
which we rewrite in the form
$$
\frac{1}{\ell}\,\ve^i{}_{mn} b^m b^n=db^i-\ve^i{}_{jk}\om^k b^j\,.
$$
Explicit calculation for $i=0,1,2$ yields:
{\jot 1.5mm
\beann
&&\om^0{_1}=0\, ,\qquad \om^1{_0}=\om^1{_2}=0\, ,
  \qquad \om^2{_1}=0\, ,                                   \\
&&\frac{2}{\ell}b^2{_0}
   =\frac{b^0{_0}'}{b^1{_1}}+\om^2{_0}
    +\frac{\om^1{_1}}{b^1{_1}}b^2{_0}\, ,
    \qquad b^2{_2}'+\om^0{_2}b^1{_1}=0\, ,                 \\
&&\frac{2}{\ell}=\frac{\om^2{_2}}{b^2{_2}}
   + \frac{\om^1{_1}}{b^1{_1}} \, ,\qquad
   \frac{2}{\ell}+\om^0{_2}\frac{b^2{_0}}{b^0{_0}b^2{_2}}
   =\frac{\om^2{_2}}{b^2{_2}}+\frac{\om^0{_0}}{b^0{_0}} \, ,\\
&& \frac{2}{\ell}-\frac{b^2{_0}'}{b^0{_0}b^1{_1}}
   =\frac{\om^1{_1}}{b^1{_1}}+\frac{\om^0{_0}}{b^0{_0}} \, .
\eeann
}
Solving these equations we find
\bea
&&\om^0=Ndx^-\, ,\qquad
  \om^1=\left(\frac{1}{\ell}+\frac{J}{r^2}\right)\frac{1}{N}dr\,,\nn\\
&&\om^2=-r\left(\frac{1}{\ell}-\frac{J}{r^2}\right)dx^-
        -\frac{J}{2r^3}dt\, .                                \lab{3.6}
\eea

This completes the derivation of the exact black hole solution in the
telaparallel geometry. The AdS vacuum solution \eq{3.2}, \eq{3.3} is
obtained for $2m=-1$, $J=0$. For other values of $m$ and $J$, the
black hole differs from the AdS vacuum, but has similar asymptotic
behaviour.

\section{Asymptotic conditions}

Dynamical structure of a field theory is determined not only by the
field equations, but also by the asymptotic conditions. An important
feature of this structure is contained in its symmetry properties. When
$\Leff<0$, the solution of \eq{2.5} possessing the maximum number of
symmetries is the AdS solution \cite{20}. It plays the role analogous
to the role of Minkowski space in the $\Leff=0$ case. Therefore, it
seems natural to choose the asymptotic behaviour in such a way that all
the dynamical variables approach the AdS configuration at large
distances. On the other hand, such an approach would exclude the
important (locally equivalent but globally distinct) black hole
geometries. Then again, these geometries are not AdS invariant --- the
minimal feature we would like to have.

Having this in mind, the concept of the {\it AdS asymptotic behaviour\/}
can be defined by imposing the following requirements \cite{1,23}:
\bitem
\item[] a) the asymptotic conditions should be invariant under the action
of the AdS group; \\
b) they should include the important black hole geometries; \\
c) the asymptotic symmetries should have well defined canonical
   generators.
\eitem
The first two requirements are studied in this section, while c) is
left for the next two sections.

\subsection{Asymptotic AdS configurations}

We begin our considerations with the point b) above. The asymptotic
behaviour of the black hole solution is easily derived from equations
\eq{3.5} and \eq{3.6}. For the triad field, it is given by
\bsubeq\lab{4.1}
\be
b^i{_\m}\sim \left( \ba{ccc}
        \dis \frac{r}{\ell}-\frac{m\ell}{r}  & 0 & 0  \\
        0 &\dis\frac{\ell}{r}+\frac{m\ell^3}{r^3} & 0\\
        \dis\frac{J}{r}  & 0       & r
                   \ea
             \right)    \, .                                \lab{4.1a}
\ee
For the simplicity of notation, the type of higher order terms on the
right hand side is not written explicitly. Similarly, the asymptotic
behaviour of the connection has the form
\be
\om^i{_\m}\sim \left( \ba{ccc}
    \dis\frac{r}{\ell^2}-\frac{m}{r} & 0
       &\dis -\frac{r}{\ell}+\frac{m\ell}{r} \\
     0 & \dis\frac{1}{r}+\frac{J\ell+m\ell^2}{r^3} & 0  \\
    \dis-\frac{r}{\ell^2}+\frac{J}{\ell r} & 0
       & \dis \frac{r}{\ell}-\frac{J}{r}
                    \ea
                \right) \, .                                \lab{4.1b}
\ee
\esubeq
The requirement b) means that the asymptotic behaviour must be such as to
allow for the black hole configuration \eq{4.1}.

In the next step, we turn to the requirement a). It can be realized by
starting with the black hole configuration \eq{4.1} and acting on it
with all possible AdS transformations. Instead of that, we shall use
the known result of such a procedure for the black hole metric, and then
transform the obtained information to the triad and connection.

The family of metrics obtained by acting on the black hole metric
\eq{3.4} with all AdS transformations, has been found by Brown and
Henneaux \cite{1}:
$$
g_{\m\n}=\left( \ba{ccc}
           \dis\frac{r^2}{\ell^2}+\cO_0  & \cO_3  & \cO_0 \\
           \cO_3 & \dis -\frac{\ell^2}{r^2}+\cO_4 & \cO_3  \\
           \cO_0 &  \cO_3  & -r^2+\cO_0
                 \ea
          \right)
\equiv   \left( \ba{ccc}
           \dis\frac{r^2}{\ell^2} & 0   & 0 \\
           0 & \dis -\frac{\ell^2}{r^2} & 0 \\
           0 & 0  & -r^2
                 \ea
          \right)+G_{\m\n}\, ,
$$
where $\cO_n$ denotes a quantity that tends to zero as $1/r^n$ or
faster, when $r\to\infty$. The set of AdS transformations is defined by
six Killing vectors (Appendix A), hence, strictly speaking, the set of
all metrics obtained from \eq{3.4} by the action of these
transformations is parametrized by six real parameters, say $\s_i$. The
meaning of the above expression for $g_{\m\n}$ is slightly different:
any $c/r^n$ term it supposed to be of the form $c(t,\vphi)/r^n$, i.e.
constants $c=c(\s_i)$ of the six parameter family are promoted to
functions $c(t,\vphi)$. This is the simplest way to characterize the
asymptotic behaviour of the family $g_{\m\n}$.

In accordance with the above result, we adopt the following asymptotic
form for the triad field:
\bsubeq\lab{4.2}
\be
b^i{_\m}= \left( \ba{ccc}
       \dis\frac{r}{\ell}+\cO_1   & O_4  & O_1  \\
       \cO_2 & \dis\frac{\ell}{r}+\cO_3  & O_2  \\
       \cO_1 & \cO_4                     & r+\cO_1
             \ea
     \right)
\equiv  \left( \ba{ccc}
       \dis\frac{r}{\ell} & 0    & 0 \\
       0 & \dis\frac{\ell}{r}    & 0 \\
       0 & 0                     & r
             \ea
     \right)+B^i{_\m}   \, .                            \lab{4.2a}
\ee
It generates the Brown--Henneaux asymptotic behaviour of the metric,
but is not uniquely determined by it. Indeed, we can apply an arbitrary
local Lorentz transformation to \eq{4.2a}, thereby changing its
asymptotics, but it will not affect the metric in any way. Our choice
of the triad asymptotics was guided by two principles: i) to obtain as
general asymptotic behaviour as possible, and ii) to evade additional
constraint relations among (otherwise arbitrary) higher order terms
$B^i{_\m}$.

Next, we use the torsion equation of motion (2.7a) to obtain the
asymptotic form of the connection:
\be
\om^i{_\m}=\left( \ba{ccc}
    \dis\frac{r}{\ell^2}+\cO_1 & \cO_2 &\dis -\frac{r}{\ell}+\cO_1 \\
    \cO_2 & \dis\frac{1}{r}+\cO_3 & \cO_2  \\
    \dis-\frac{r}{\ell^2}+\cO_1 & \cO_2 & \dis \frac{r}{\ell}+\cO_1
                    \ea
                \right)
\equiv  \left( \ba{ccc}
    \dis\frac{r}{\ell^2} & 0 &\dis -\frac{r}{\ell} \\
    0 & \dis\frac{1}{r}  & 0  \\
    \dis-\frac{r}{\ell^2} & 0 & \dis \frac{r}{\ell}
                    \ea
                \right)+\Om^i{_\m} \, .                  \lab{4.2b}
\ee
\esubeq
Again, the higher order terms $\Om^i{_\m}$ are considered arbitrary and
independent of those in \eq{4.2a}. One can check that the asymptotic
conditions \eq{4.2} are indeed invariant under the action of the AdS
group.

\subsection{Asymptotic symmetries}

We are now going to examine the symmetries of the asymptotic conditions
\eq{4.2}. The parameters of gauge transformations that leave the
conditions \eq{4.2} invariant are determined by the relations
\bsubeq\lab{4.3}
\bea &&\ve^{ijk}\th_j
b_{k\m}-\xi^{\r}{}_{,\,\m}b^i{_\r}
        -\xi^{\r}\pd_\r b^i{}_{\m}=\d_0 B^i{_\m}\, ,     \lab{4.3a}\\
&&-\th^i{}_{,\,\m}+\ve^{ijk}\th_j\om_{k\m}
        -\xi^{\r}{}_{,\,\m}\om^i{_\r}
        -\xi^{\r}\pd_\r\om^i{}_{\m}=\d_0\Om^i{_\m}\, .   \lab{4.3b}
\eea
\esubeq
Acting on a specific field satisfying the adopted asymptotic conditions,
these transformations change the form of the non-leading terms
$B^i{_\m}$, $\Om^i{_\m}$. One should stress that the symmetry
transformations defined in this way differ from the usual asymptotic
symmetries, which act according to the rule $\d_0 b^i{_\m}=0$,
$\d_0\om^i{_\m}=0$.

We shall find the gauge parameters in three steps.

\prg{1.} The symmetric part of the first equation multiplied by
$b_{i\n}$ (six relations) has the form
\be
-\xi^{\r}{}_{,\m}g_{\n\r} -\xi^{\r}{}_{,\n}g_{\m\r}
-\xi^{\r}\pd_\r g_{\m\n}=\d_0 G_{\m\n} \, ,               \lab{4.4}
\ee
which defines the transformation rule of the metric. The matrix
$\d_0 G_{\m\n}$ has the same form as $G_{\m\n}$. If we define the
expansion of $\xi^\m$ in powers of $r^{-1}$,
$$
\xi^\m=\sum_{n=-1}^{\infty}\xi^\m_n r^{-n}\, ,
$$
the condition \eq{4.4} yields
\bea
&&\xi^0_{-1}=\xi^0_1=\xi^0_3=0\, ,\qquad \xi^1_0=0\, ,
  \qquad \xi^2_{-1}=\xi^2_1=\xi^2_3=0\, , \nn\\
&& \xi^0_2=\frac{\ell^4}{2}\,\xi^0_{0,00}\, ,\qquad
   \xi^1_{-1}=-\xi^0_{0,0}\, ,\qquad
   \xi^2_2=-\frac{\ell^2}{2}\,\xi^2_{0,22}\, ,        \nn\\
&& \xi^2_{0,2}=\xi^0_{0,0}\, ,\qquad
   \xi^2_{0,0}=\frac{1}{\ell^2}\,\xi^0_{0,2}\, . \nn
\eea
Then, after introducing the notation
$$
\xi^0_0=\ell T(t,\vphi)\, ,\qquad \xi^2_0=S(t,\vphi)\, ,
$$
the solution of the above equations takes the form
\bsubeq\lab{4.5}
\bea
&&\xi^0=\ell\left[ T
  +\frac{1}{2}\left(\frac{\pd^2 T}{\pd t^2}\right)
              \frac{\ell^4}{r^2}\right] +\cO_4\, ,      \\
&&\xi^2=S-\frac{1}{2}\left(\frac{\pd^2 S}{\pd\vphi^2}\right)
              \frac{\ell^2}{r^2}+\cO_4\, ,              \\
&&\xi^1=-\ell\left(\frac{\pd T}{\pd t}\right)r+\cO_1\, ,
\eea
where the functions $T(t,\vphi)$ and $S(t,\vphi)$ satisfy the
conditions
$$
\frac{\pd T}{\pd\vphi}=\ell\frac{\pd S}{\pd t}\, ,\qquad
\frac{\pd S}{\pd\vphi}=\ell\frac{\pd T}{\pd t}\, .       \eqno(4.6)
$$

The above equations define the conformal group of transformations at
large distances \cite{1}. Whether this group will survive as the
asymptotic symmetry of our teleparallel theory depends on the remaining
conditions in \eq{4.3}.

\prg{2.} After having used six components of \eq{4.3a} to find the form
of $\xi^\m$, we shall now determine $\th^i$ from the remaining three
components. They yield the relations
\bea
&&\th^1 b^2{_1}-\th^2 b^1{_1}-\xi^0{}_{,1}b^0{_0}=\cO_4 \, , \qquad
  \th^1 b^2{_2}-\xi^0{}_{,2}b^0{_0}=\cO_1 \, ,\nn\\
&&\th^0 b^2{_2}-\xi^\r{}_{,2}b^1{_\r}=\cO_2 \, ,\nn
\eea
with the solution
\bea
&&\th^0=-\frac{\ell^2}{r}T_{,02}+\cO_3\, ,\nn\\
&&\th^2=\frac{\ell^3}{r}T_{,00}+\cO_3\, , \nn\\
&&\th^1=T_{,2}+\cO_2\, .                                    \lab{4.5d}
\eea
\esubeq
\setcounter{equation}{6}

\prg{3.} We have seen that the invariance conditions \eq{4.3a}
completely define the parameters $\xi^\m$ and $\th^i$, as shown in
\eq{4.5}. Now, the last thing to check is whether the transformation
law for the connection leads to any new limitation on the parameters.
Explicit calculation shows that the connection configuration \eq{4.2b}
is also invariant under the transformations defined by the above
parameters.

The allowed form of the parameters $T$ and $S$ is obtained by solving
equations (4.6). Rewriting these conditions in the form
$$
\pd_\pm(T \mp S)=0 \, ,
$$
with $2\pd_\pm=\ell\pd_0 \pm\pd_2$, we find that the general solution
is given by
\be
T+S=f(x^+)\, ,\qquad  T-S=g(x^-)\, ,                         \lab{4.7}
\ee
where $f$ and $g$ are two arbitrary, periodic functions.

We have found that our gauge parameters $(\xi^\m,\th^i)$ must be of the
form \eq{4.5}, in order to preserve the adopted boundary conditions
\eq{4.2}. At large distances, the parameters $(\xi^\m,\th^i)$ are
determined by the functions $(T,S)$ which define the conformal
symmetry at the boundary of our teleparallel spacetime. (This will be
shown in Sec. VI, while the related analysis in Riemannian case is
given in Ref. \cite{1}.) However, the complete gauge group defined in
this way contains also the {\it residual\/} (or pure) gauge
transformations, characterized by the higher $\cO_n$ terms in \eq{4.5},
the only terms that remain after imposing $T=S=0$. As we shall see, the
residual gauge transformations do not contribute to the values of the
conserved charges, and consequently, their generators vanish weakly. In
order to ``eliminate" the residual gauge transformations from our
discussion, we introduce the concept of {\it asymptotic symmetry\/} in
the following way:
\bitem
\item[] the asymptotic symmetry group is defined as the factor group
of the gauge group determined by \eq{4.5}, with respect to the residual
gauge group.
\eitem
In other words, two asymptotic symmetry transformations are identified
if their parameters $(\xi^\m, \th^i)$ have identical $(T,S)$ pairs,
while any difference stemming from the higher $\cO_n$ terms in \eq{4.5}
is ignored.

\prg{}
In conclusion, the adopted asymptotic behaviour \eq{4.2} defines the
most general configuration space of the theory that respects our
requirements a) and b) formulated at the beginning of this section.
The related symmetry structure is determined by the parameters \eq{4.5}.
In order to verify the status of the last requirement c), it is
necessary to explore the canonical structure of the theory.

\section{Hamiltonian structure}

Gauge symmetries of a dynamical system are best described by the
canonical generators. After clarifying the canonical and gauge structure
of our theory of gravity, we shall be able to better understand the
meaning of the adopted asymptotic conditions.

The action of the theory \eq{2.9} can equivalently be written as
\be
I= a\int d^3x\ve^{\r\m\n}\left[  b^i{_\r}\left(R_{i\m\n}
   -\frac{1}{\ell}T_{i\m\n}\right) +\frac{4}{3\ell^2}\,
    \ve_{ijk}b^i{_\r}b^j{_\m}b^k{_\n}\right]\, .            \lab{5.1}
\ee

\subsection{Hamiltonian and constraints}

Starting from the definition of the the canonical momenta
$(\pi_i{^\m},\Pi_i{^\m})$, corresponding to the basic Lagrangian
variables $(b^i{_\m},\om^i{_\m})$, we find the following primary
constraints:
\bea
&&\phi_i{^0}\equiv\pi_i{^0}\approx 0\, ,\qquad
  \Phi_i{^0}\equiv\Pi_i{^0}\approx 0\, ,                     \nn\\
&&\phi_i{^\a}\equiv\pi_i{^\a}
  +\frac{2a}{\ell}\ve^{0\a\b}b_{i\b}\approx 0\, ,\qquad
  \Phi_i{^\a}\equiv\Pi_i{^\a}-2a\ve^{0\a\b}b_{i\b}\approx 0\,.\lab{5.2}
\eea

Since the Lagrangian is linear in velocities, the canonical Hamiltonian
is determined by the formula $\cH_c=-\cL(\dot b^i{_\m}=\dot\om^i{_\m}=0)$.
It is linear in unphysical variables $(b^i{_0},\om^i{_0})$, as we expect:
$$
\cH_c=b^i{_0}\cH_i+\om^i{_0}\cK_i+\pd_\a D^\a\, ,
$$
where
\bea
&&\cH_i=-a\ve^{0\a\b}\left(R_{i\a\b}-\frac{2}{\ell}T_{i\a\b}
          +\frac{4}{\ell^2}\,\ve_{ijk}b^j{_\a}b^k{_\b}\right)\, ,\nn\\
&&\cK_i=-a\ve^{0\a\b}\left(T_{i\a\b}
          -\frac{2}{\ell}\ve_{ijk}b^j{_\a}b^k{_\b}\right)\, ,\nn\\
&&D^\a=2a\ve^{0\a\b}b^i{_\b}\left(\om_{i0}
                                -\frac{1}{\ell}b_{i0}\right)\, .\nn
\eea
The total Hamiltonian has the form
\be
\cH_T=b^i{_0}\cH_i+\om^i{_0}\cK_i
  +u^i{_\m}\phi_i{^\m}+ v^i{_\m}\Phi_i{^\m}+\pd_\a D^\a\, . \lab{5.3}
\ee

The consistency conditions of the sure primary constraints
$\pi_i{^0}$ and $\Pi_i{^0}$ lead to the secondary constraints:
\be
\cH_i\approx 0\, ,\qquad \cK_i\approx 0\, ,                 \lab{5.4}
\ee
which can be equivalently written as
$$
R_{i\a\b}\approx 0\, ,\qquad
T_{i\a\b}\approx\frac{2}{\ell}\ve_{imn}b^m{_\a}b^n{_\b}\, .
$$

The consistency of the remaining primary constraints $\phi_i{^\a}$ and
$\Phi_i{^\a}$ leads to the determination of the multipliers $u^i{_\b}$
and $v^i{_\b}$:
\bea
&& u^i{_\b}+\ve^{imn}\om_{m0}b_{n\b}-\nabla_\b b^i{_0}
           =\frac{2}{\ell}\ve^{imn}b_{m0}b_{n\b}\, , \nn\\
&& v^i{_\b}-\nabla_\b\om^i{_0}=0 \, .                      \lab{5.5}
\eea
Since the equation of motion for $b^i{_\b}$ has the form $u^i{_\b}=\dot
b^i{_\b}$, the first relation can be written as the field equation
$T^i{}_{0\b}=(2/\ell)\ve^{imn}b_{m0}b_{n\b}$, with $\dot b^i{_\b}\to
u^i{_\b}$. Similarly, the second relation is on shell equivalent to the
field equation $R^i{}_{0\b}=0$, with $\dot\om^i{_\b}\to v^i{_\b}$. The
result is obtained using the following Poisson brackets (PBs) involving
$\phi_i{^\a}$ and $\Phi_i{^\a}$:
\bea
&&\{\phi_i{^\a},\phi_j{^\b}\}
  =\frac{4a}{\ell}\ve^{0\a\b}\eta_{ij}\d\, ,           \nn\\
&&\{\phi_i{^\a},\Phi_j{^\b}\}=-2a\ve^{0\a\b}\eta_{ij}\d\, ,
   \qquad  \{\Phi_i{^\a},\Phi_j{^\b}\}=0\, ,           \nn
\eea
and
\bea
&&\{\phi_i{^\a},\cH_j\}=-\frac{2}{\ell}\{\phi_i{^\a},\cK_j\}
                     =-\frac{2}{\ell}\{\Phi_i{^\a},\cH_j \} \, ,\nn\\
&&\{\Phi_i{^\a},\cH_j\} =2a\ve^{0\a\b}\left[\eta_{ij}\pd_\b\d
                      -\ve_{ijn}\left(\om^n{_\b}
                      -\frac{2}{\ell}b^n{_\b}\right)\d\right]\, ,\nn\\
&&\{\Phi_i{^\a},\cK_j\}=-2a\ve^{0\a\b}\ve_{ijn}b^n{_\b}\d\, . \nn
\eea
Here, we use the notation $\{A,B\}=\{A(x),B(y)\}$, $\d=\d(x-y)$,
$\pd_\b=\pd/\pd x^\b$.

Replacing the expressions \eq{5.5} into the total Hamiltonian \eq{5.3},
we obtain
\bsubeq\lab{5.6}
\bea
&&\cH_T=\hcH_T +\pd_\a\bD^\a\, ,\nn\\
&&\hcH_T=b^i{_0}\bcH_i+\om^i{_0}\bcK_i
         +u^i{_0}\pi_i{^0}+v^i{_0}\Pi_i{^0} \, ,           \lab{5.6a}
\eea
where
\bea
&&\bcH_i=\cH_i-\nabla_\b\phi_i{^\b}
              +\frac{2}{\ell}\ve_{imn}b^m{_\b}\phi^{n\b}\, , \nn\\
&&\bcK_i=\cK_i-\nabla_\b\Phi_i{^\b}
              -\ve_{imn}b^m{_\b}\phi^{n\b}\, ,              \nn\\
&&\bD^\a=D^\a+b^k{_0}\phi_k{^\a}+\om^k{_0}\Phi_k{^\a}\, .   \lab{5.6b}
\eea
\esubeq

Further investigation of the consistency requirements is facilitated by
observing that the secondary constraints $\bcH_i,\bcK_i$ obey the
following PB algebra:
\bea
&&\{\bcH_i,\bcH_j\}=\frac{2}{\ell}\ve_{ijk}\bcH^k\d\, ,\qquad
  \{\bcH_i,\bcK_j\}=-\ve_{ijk}\bcH^k\d\, ,\nn\\
&&\{\bcK_i,\bcK_j\}=-\ve_{ijk}\bcK^k\d\, .                  \lab{5.7}
\eea
Indeed, we can now conclude that consistency conditions of the secondary
constraints are identically satisfied, which completes the consistency
procedure.

The complete dynamical classification of the constraints is given in
Table 1.
\begin{center}
\doublerulesep .5pt
\begin{tabular}{|l|l|l|}
\multicolumn{3}{l}{Table 1. Classification of constraints}\\
                                                       \hline\hline
\rule{0pt}{12pt}&~first class\phantom{x}&~second class\phantom{x}\\
                                                       \hline
\rule[-1pt]{0pt}{15pt}
primary &~$\p_i{^0},\Pi_i{^0}$ &~$\phi_i{^\a},\Phi_i{}^{\a}$ \\
                                                       \hline
\rule[-1pt]{0pt}{15pt}
secondary\phantom{x} &~$\bcH_i,\bcK_i$     &~               \\
                                                       \hline\hline
\end{tabular}
\end{center}\bigskip
We display here, for completeness, the PBs between
$(\phi_i{^\a},\Phi_i{^\a})$ and $(\bcH_j,\bcK_j)$:
\bea
&&\{\phi_i{^\a},\bcH_j\}=\frac{2}{\ell}\,\ve_{ijk}\phi^{k\a}\d\, ,
  \qquad   \{\phi_i{^\a},\bcK_j\}=-\ve_{ijk}\phi^{k\a}\d\, , \nn\\
&&\{\Phi_i{^\a},\bcH_j\}=-\ve_{ijk}\phi^{k\a}\d\, ,\qquad
  \{\Phi_i{^\a},\bcK_j\}=-\ve_{ijk}\Phi^{k\a}\d\, .          \nn
\eea

\subsection{Gauge generators}

The presence of arbitrary multipliers in the total Hamiltonian
indicates the existence of gauge symmetries in the theory. The
canonical gauge generators can be constructed using the general method
of Castellani \cite{24}.

Starting from the primary first class constraints $\pi_i{^0}$ and
$\Pi_i{^0}$, we find the form of the respective gauge generators:
\bea
&&G[\eps]=\dot\eps^i\pi_i{^0}
      +\eps^i\left[\,\bcH_i - \ve_{ijk}\left(\om^j{_0}
      -\frac{2}{\ell}\,b^j{_0}\right)\pi^{k0}\,\right] \, , \nn\\
&&G[\t]=\dot\t^i\Pi_i{^0}
    +\t^i\left[\,\bcK_i-\ve_{ijk}\left( b^j{_0}\,\pi^{k0}
    +\om^j{_0}\,\Pi^{k0}\right)\,\right]\, .              \lab{5.8}
\eea
The complete gauge generator is given as the sum $G=G[\eps]+G[\t]$, and
it produces the following gauge transformations on the fields
($\d_0\phi\equiv \{\phi\,,G\}$):
\bea
&&\d_0\,b^i{_{\m}}=\nabla_{\m}\eps^i
  -\frac{2}{\ell}\,\ve^i{}_{jk}b^j{_{\m}}\eps^k
  +\ve^i{}_{jk}b^j{_{\m}}\tau^k\, ,                      \nn\\
&&\d_0\,\om^i{}_{\m}=\nabla_{\m}\tau^i \,.               \nn
\eea
Introducing now the new parameters $\xi^{\m}$ and $\th^i$,
\be
\eps^i=-\xi^{\m}b^i{}_{\m}\, ,\qquad
\t^i=-(\th^i+\xi^\m\om^i{_\m})\, ,                          \lab{5.9}
\ee
the transformation law  takes the form
\bea
&&\d_0b^i{_{\m}}= \ve^{ijk}\th_jb_{k\m}
  -\xi^{\r}{}_{,\,\m}b^i{}_{\r}-\xi^{\r}b^i{}_{\m,\,\r}
  -\xi^\r\left(T^i{}_{\m\r}
          -\frac{2}{\ell}\,\ve^{ijk}b_{j\m}b_{k\r}\right)\,, \nn\\
&&\d_0\om^i{_{\m}}= -\nabla_\m\th^i
  -\xi^{\r}{}_{,\,\m}\om^i{}_{\r}-\xi^{\r}\om^i{}_{\m,\,\r}
  -\xi^\r R_{i\m\r}\, ,\nn
\eea
which is in complete agreement with the transformations \eq{2.1a} {\it
on-shell\/}. The gauge generator $G$, expressed in terms of the new
parameters $\xi^\m$ and $\th^i$, is given by
\bea
&& G=-G_1-G_2\, ,\nn\\
&&G_1\equiv\dot\xi^\r\left(b^i{_\r}\pi_i{^0}+\om^i{_\r}\Pi_i{^0}\right)
    +\xi^\r\left[b^i{_\r}\bcH_i +\om^i{_\r}\bcK_i
    +(\pd_\r b^i{_0})\pi_i{^0}+(\pd_\r\om^i{_0})\Pi_i{^0}\right]\,,\nn\\
&&G_2\equiv\dot\th^i\Pi_i{^0}
    +\th^i\left[\bcK_i-\ve_{ijk}\left( b^j{_0}\pi^{k0}
    +\om^j{_0}\,\Pi^{k0}\right)\right]\, .                  \lab{5.10}
\eea
Here, the time derivatives $\dot b^i{_\m}$ and $\dot\om^i{_\m}$ are
shorts for $u^i{_\m}$ and $v^i{_\m}$, respectively.
The result is obtained by discarding terms that produce trivial
transformations on-shell. Note that the time translation generator
$G_1[\xi^0]$ is defined in terms of $\hcH_T$:
$$
G_1[\xi^0]=\dot\xi^0\left(b^i{_0}\pi_i{^0}+\om^i{_0}\Pi_i{^0}\right)
           +\xi^0\hcH_T \, .
$$
(In the above expressions for the gauge generators, we did not write
the integration symbol $\int d^2x$ in order to simplify the
notation. Later, where necessary, the integration symbol will be
restored.)

To complete the analysis of the asymptotic structure of phase space, we
shall now define the behaviour of momentum variables at large
distances. Our procedure is based on the following general principle:
the expressions that vanish on-shell should have an {\it arbitrarily
fast asymptotic decrease}, as no solution of the field equations is
thereby lost. Applied to the primary constraints of the theory, this
principle gives us the asymptotic behaviour of the momenta $\p_i{^\m}$
and $\Pi_i{^\m}$:
\bea
&&\pi_i{^0}= \hcO\, ,\qquad
  \pi_i{^\a}=-\frac{2a}{\ell}\ve^{0\a\b}b_{i\b}+\hcO\,,    \nn\\
&&\Pi_i{^0}=\hcO\, ,\qquad
  \Pi_i{^\a}=2a\,\ve^{0\a\b}b_{i\b}+\hcO\, ,               \lab{5.11}
\eea
where $\hcO$ denotes a term with arbitrarily fast decrease.
The asymptotic form of the secondary constraints, as well as some of
the equations of motion, are given in Appendix B. They further
refine the asymptotic behaviour of phase space variables, and serve
as a tool to prove the finiteness of our improved gauge generators.

We are now ready to discuss the impact of the adopted boundary
conditions on the form of the canonical generator.

\section{Asymptotic symmetry}

After having introduced the notion of the asymptotic symmetry group in
section IV, we now wish to continue with the related canonical
analysis. For the asymptotic generator, we use the simple notation
$G[T,S]$, indicating the irrelevance of the residual gauge parameters
in \eq{4.5}. (Although the concept of an asymptotic generator is
different from the gauge generator in general, we use the same symbol
$G$ for both, in order to keep the notation simple.)

In what follows, we shall construct the improved form of the general
gauge generator $G$, assuming the asymptotic conditions \eq{4.2} and
\eq{4.5}, and prove the conservation of the corresponding charges;
then, we shall investigate the form and properties of the canonical
algebra of asymptotic generators, and find the value of the related
central charge.

\subsection{Boundary terms}

In the Hamiltonian theory, the generators of symmetry transformations
act on dynamical variables via the PB operation, which is defined in
terms of functional derivatives. A functional
$F[\vphi,\p]=\int d^3x f(\vphi,\pd_\m\vphi,\p,\pd_\n\p)$
has well defined functional derivatives if its variation can be
written in the form
$$
\d F=\int d^3x\bigl[ A(x)\d\vphi(x)+B(x)\d\p(x)\bigr]\,,
$$
where terms $\d\vphi_{,\m}$ and $\d\p_{,\n}$ are absent. In
addition, the well defined phase space functionals have to be
finite.

Our general gauge generator $G$ does not meet these requirements,
although when acting on local expressions, it produces the correct
transformation laws. We shall now improve its form by adding an
appropriate surface term, so that it can also act on global
phase space functionals \cite{18}.

We begin by considering the variation of $G_2$:
\bea
\d G_2&=&\th^i\d\bcK_i+R=\th^i\d\cK_i+\pd\hcO+R     \nn\\
  &=&-2a\ve^{\a\b 0}\th^i\pd_\a\d b_{i\b}+\pd\hcO+R       \nn\\
   &=&-2a\ve^{\a\b 0}\pd_\a\left(\th^i\d b_{i\b}\right)+\pd\hcO+R
    =\pd\cO_2+R \, .                                     \nn
\eea
The last equality is a consequence of the relation $\th^i\d
b_{i\b}=\cO_2$, derived from the asymptotic conditions \eq{4.5d} and
\eq{4.2}. The total divergence term $\pd\cO_2$ gives a vanishing
contribution after integration, as follows from the Stokes theorem:
$$
\int_{\cM_2}d^2x\,\pd_\a v^\a=\int_{\pd\cM_2}v^\a df_\a
                               =\int_0^{2\pi} v^1 d\vphi
\qquad (df_\a\equiv\ve_{\a\b}dx^\b)\, .
$$
In the last equality, the boundary of $\cM_2$ is taken to be the
circle at infinity, parametrized by the angular coordinate $\vphi$.
Thus, $G_2$ is a regular generator for which there is no need to
introduce any boundary term:
\be
{\rm Boundary~term~for~} G_2 =0\, .                        \lab{6.1}
\ee

The variation of $G_1$ yields
\bea
\d G_1&=&\xi^\r\left(b^i{_\r}\d\bcH_i
                  +\om^i{_\r}\d\bcK_i\right)+\pd\hcO+R        \nn\\
  &=&-2a\ve^{\a\b 0}\pd_\a\left[
      \xi^\r b^i{_\r}\d\left(\om_{i\b}-\frac{2}{\ell}b_{i\b}\right)
     +\xi^\r\om^i{_\r}\d b_{i\b}\right] + \pd\hcO + R \, .    \nn
\eea
We shall now focus on the terms containing $\xi^0$, $\xi^1$ and
$\xi^2$.

\prg{1.} Using the preceding  result and the asymptotic conditions
\eq{4.5}, \eq{4.2},  the variation of the generator $G_1[\xi^0]$ is
shown to have the form
$$
\d G_1[\xi^0]=-2a\ve^{\a\b 0}\pd_\a\left[
      \xi^0 b^0{_0}\d\left(\om^0{_\b}-\frac{1}{\ell}b^0{_\b}
      +\frac{1}{\ell}b^2{_\b}\right)\right] + \pd\cO_2 + R \, ,
$$
wherefrom
\bea
&&\d G_1[\xi^0] = -\d\pd_\a\left(\xi^0 \cE^{\a}\right)
                  + \pd\cO_2+R \,,                           \nn\\
&&\cE^{\a} = 2a\,\ve^{\a\b0}\left(\om^0{_\b}
                  +\frac{1}{\ell}\,b^2{_\b}
                  -\frac{1}{\ell}\,b^0{_\b}\right)b^0{_0}\,.  \nn
\eea
The improved time translation generator reads:
\bsubeq\lab{6.2}
\bea
&&\tG_1[\xi^0]= G_1[\xi^0] + E[\xi^0]\, ,\nn\\
&&E[\xi^0]=\oint\xi^0\cE^{\a}df_{\a}
          =\int_0^{2\pi}d\vphi\,\xi^0\cE^1\, .             \lab{6.2a}
\eea
From $\xi^0=\cO_0$, $\cE^1=\cO_0$, it follows that $E$ must be finite.

Note that, for $\xi^0=1$, the generator $G_1$ reduces to $\hcH_T$,
so that the corresponding boundary term defines the improved
Hamiltonian, and has the meaning of energy:
\bea
&& \tH_T = \hat H_T + E_0 \,, \nn\\
&& E_0=\oint\cE^{\a}df_{\a}=\int_0^{2\p}\!\cE^1d\vphi\,.  \lab{6.2b}
\eea
\esubeq

\prg{2.} In a similar way, we find that the variation of $G_1[\xi^2]$
has the form
$$
\d G_1[\xi^2]=2a\,\ve^{\a\b 0}\pd_\a\left[
   \xi^2 b^2{_2}\d\left(\om^2{_\b}+\frac{1}{\ell}b^0{_\b}
  -\frac{1}{\ell}b^2{_\b}\right)\right]+\pd\cO_2 + R \,,
$$
whereupon we conclude that
\bea
&&\d G_1[\xi^2] = -\d\pd_\a\left(\xi^2 \cM^{\a}\right)
                       + \pd\cO_2+R \, ,                    \nn\\
&&\cM^{\a}=-2a\,\ve^{\a\b 0}\left(\om^2{_\b}+\frac{1}{\ell}\,b^0{_\b}
             -\frac{1}{\ell}\,b^2{_\b}\right)b^2{_2}\, .    \nn
\eea
The improved spatial rotation generator reads:
\bsubeq
\bea
&&\tG_1[\xi^2]= G_1[\xi^2] + M[\xi^2]\, ,\nn\\
&&M[\xi^2]=\oint\xi^2\cM^{\a}df_{\a}
          =\int_0^{2\pi}d\vphi\,\xi^2\cM^1 \,.            \lab{6.3a}
\eea
The boundary term
\be
M_0=\oint{\cal M}^{\a}df_{\a}=\int_0^{2\pi}\!\cM^1 d\vphi \lab{6.3b}
\ee
\esubeq
represents the angular momentum of the system. The adopted
asymptotics ensures the finiteness of the improved generator.

\prg{3.} Analogous considerations in the case of $G_1[\xi^1]$ lead
us to the conclusion that this generator is regular,
\be
{\rm Boundary~term~for~} G_1[\xi^1]=0\, .                  \lab{6.4}
\ee
Therefore, the improved gauge generator $\tG$ is given by the
expression
\bea
&&\tG= G +\G \, ,                                         \nn\\
&&\G=-\oint df_\a\Big( \xi^0\cE^{\a}+\xi^2\cM^{\a}\Big)
  =-\int_0^{2\p}\!d\vphi\Big( \ell T\cE^1+S\cM^1\Big)\,.   \lab{6.5}
\eea
The adopted asymptotic behaviour of fields and parameters guarantees
finiteness of the surface term $\G$. Being a linear combination of
constraints, the generator $G$ is finite itself, and therefore, the
improved generator $\tG$ is well defined, differentiable functional
on its whole domain.

As we can see, the surface term $\G$ depends only on the parameters
$(T,S)$, and not on the higher order terms in \eq{4.5}. Thus, it is
only the {\it asymptotic} generators that have non-trivial boundary
terms, and, consequently, do not vanish weakly. We expect the
corresponding conserved charges to be physically non-trivial. On the
other hand, the {\it residual\/} gauge generators are characterized by
vanishing $\G$, and can only have zero charges \cite{1,22}.

\subsection{Canonical algebra}

We now wish to find the form of the canonical algebra of the improved
asymptotic generators, which contains important information on the
symmetry structure of the asymptotic dynamics. We will use this algebra
to prove the conservation of the boundary terms. Introducing the notation
$$
G'\equiv G[T',S']\,,\qquad G''\equiv G[T'',S'']\,,\ \ \ldots\ ,
$$
(similarly $\,\d_0'\equiv \d_0[T',S']\,$, etc.) we find that
\be
\left\{\tG'',\,\tG'\right\}=\d_0'\tG''\approx\d_0'\G'' \,, \lab{6.6}
\ee because every symmetry generator commutes with all the
constraints of the theory. The right-hand side of the above equation
represents the transformation of the surface term $\G''$ under the
action of the generator $\tG'$. Using the transformation rules
\eq{2.1a} with parameters \eq{4.5}, and refined asymptotic conditions
\eq{B3a} of Appendix B, we find:
\bea
&&\d_0(\ell\cE^1)=
  -\cM^1T_{,\,2}-\ell \cE^1S_{,\,2}
  -\left(\cM^1T+\ell \cE^1S\right)_{,\,2}
  + 2a\ell\,S_{,\,222}+\cO_2 \,,                           \nn\\
&&\d_0\cM^1=
  -\ell\cE^1T_{,\,2}-\cM^1S_{,\,2}
  -\left(\ell\cE^1T+\cM^1S\right)_{,\,2}
  +2a\ell\,T_{,\,222}+\cO_2 \,.                            \lab{6.7}
\eea
The above result implies
\be
\d_0'\G''=\G'''+C''' \,,                                   \lab{6.8}
\ee
where the parameters $T'''$, $S'''$  are determined by the relations
\bea
&&T'''=T'S''_{,\,2}-T''S'_{,\,2}
       +S'T''_{,\,2}-S''T'_{,\,2} \, ,                    \nn\\
&&S'''=S'S''_{,\,2}-S''S'_{,\,2}
       +T'T''_{,\,2}-T''T'_{,\,2}\, ,                     \lab{6.9}
\eea
and $C'''\equiv C[T',S'\,;\,T'',S'']$ is the {\it central term\/} of
the canonical algebra:
\be
C'''=2a\ell\int d\vphi\left(S''_{,\,2}T'_{,\,22}-
    S'_{,\,2}T''_{,\,22}\right)\,.                       \lab{6.10}
\ee
Combining Eqs. \eq{6.6} and \eq{6.8}, one finds that the PB of the
asymptotic symmetry generators has the form
$\{\tG'',\,\tG'\}\approx \G''' + C'''$, which implies the {\it weak\/}
equality
\be
\left\{\tG'',\,\tG'\right\}\approx \tG''' + C''' \,.     \lab{6.11}
\ee

It is known that the PB of two well defined generators must also be a
well defined generator \cite{25}. To promote \eq{6.11} to the strong
equality, we have to prove that there are no well defined asymptotic
generators that weakly vanish (on the space of all solutions). Indeed,
it has been shown in Appendix C that every asymptotic generator $\tG$
has a non-trivial surface term. Therefore, there holds the strong
equality
\be
\left\{\tG'',\,\tG'\right\} =\tG''' + C''' \,.           \lab{6.12}
\ee

\subsection{Conservation laws}

Let us first note that the improved total Hamiltonian $\tH_T$ is one
of the generators $\tG[T,\,S]$. Indeed, the choice $T=1$, $S=0$
in \eq{6.5} gives
\be
\tG[1,0]=-\ell\,\tH_T \,.                                \lab{6.13}
\ee
As a consequence, the commutator of the generator $\tG$ with
the improved Hamiltonian $\tH_T$ {\it does not contain central
term\/}:
\be
\left\{\tG[T,S],\,\tH_T\right\}=
-\frac{1}{\ell}\left\{\tG[T,S],\,\tG[1,0]\right\}
=-\frac{1}{\ell}\,\tG[\,S_{,\,2}\,,T_{,\,2}]\, .         \lab{6.14}
\ee
The last term in the above equation is obtained by observing that
$(T',S')=(1,0)$, $(T'',S'')=(T,S)$ implies $C'''=0\,$,
$(T''',S''')=(S_{,2}\,,T_{,2})$. Now, with the help of relations (4.6),
we find
\bea
\frac{d}{dt}\tG[T,S]&=&\frac{\pd\tG}{\pd t}+\left\{\tG,\,\tH_T\right\}
\approx \frac{\pd\G}{\pd t}
        -\frac{1}{\ell}\,\G[S_{,\,2}\,,\,T_{,\,2}]       \nn\\
&=&-\int d\vphi \left(\cM^1S_{,\,0}+\ell\,\cE^1T_{,\,0}\right)
        +\frac{1}{\ell}\int d\vphi\left(\cM^1T_{,\,2}
        +\ell\,\cE^1S_{,\,2}\right)=0\, .                \nn
\eea
Thus, the asymptotic generator $\tG[T,S]$ is conserved for every
allowed choice of the parameters $T$, $S$. This also implies the
conservation of the boundary term $\G$, as $\tG\approx \G$, and the
constraints of the theory are conserved anyway:
\be
\frac{d}{dt}\,\G[T,S]\approx 0 \,.                        \lab{6.15}
\ee

To test the obtained result, we shall calculate the value of all the
conserved charges for the BTZ black hole solution \eq{3.5}, \eq{3.6}.

First, note that the black hole solution depends on the radial
coordinate only, and consequently, the terms $\cE^1$ and $\cM^1$ in the
surface integral $\G$ behave as constants. Second, the parameters $T$,
$S$ are periodic functions, as given in \eq{4.7}. This means that only
{\it zero modes} in the Fourier decomposition of $T$, $S$ survive the
integration in $\G$. Thus, there are only two independent non-vanishing
charges for the black hole solution, and they are given by two
inequivalent choices of {\it constants\/} $T$ and $S$. If we take, let us
say,  $T=1$, $S=0$ as the first choice, and $T=0$, $S=1$ as the second
one, all the other non-zero charges will be given as linear
combinations of these two.

The particular choice $T=1$, $S=0$ gives $\G[1,0]=-\ell E_0\,$, which
means that the corresponding conserved charge is the energy $E_0$. Its
value for the black hole solution is found to be $E_0=4\p am$, but
taking into account that $a=1/16\p G=1/4\pi$ (in units $4G=1$),
we obtain
$$
E_0({\rm black~hole}) = m\,.
$$
The second choice $T=0$, $S=1$, on the other hand, leads to
$\G[0,1]=-M_0\,$. The corresponding conserved charge is the angular
momentum $M_0$, and its black hole value reads
$$
M_0({\rm black~hole}) = J\,.
$$
We see that constants $m$ and $J$, which parametrize the black hole
solution, have the meaning of energy and angular momentum,
respectively. We also see that there are no other independent conserved
charges. Geometrically, these two charges parametrize globally
inequivalent asymptotically AdS spaces \cite{22}.

\subsection{Central charge}

The canonical algebra \eq{6.12} can be brought to a more recognizable
form by using the representation in terms of Fourier modes. The
solutions \eq{4.7} for the parameters $T$ and $S$ are then written in
the form:
\bea
&& T=\sum_{-\infty}^{\infty}
     \left( a_ne^{inx^+}+\bar a_ne^{inx^-}\right)\,,\nn\\
&& S=\sum_{-\infty}^{\infty}
     \left( a_ne^{inx^+}-\bar a_ne^{inx^-}\right)\,,     \lab{6.16}
\eea
where $x^{\pm}=(t/\ell)\pm\vphi$, and the reality of $T,S$ implies
$a_{-n}=a_n^*$, $\bar a_{-n}=\bar a_n^*$. The asymptotic generator
$\tG[T,S]$ is a linear, homogeneous function of its parameters $T$ and
$S$, since the asymptotic symmetry is defined up to the residual gauge
transformations generated by $\tG[T=S=0]$. Therefore, using the above
Fourier decomposition, we can write
\be
\tG[T,S]=-2\sum_{-\infty}^{\infty}
         \left(a_nL_n+\bar a_n\bL_n\right)\,.             \lab{6.17}
\ee
The new asymptotic generators $L_n$, $\bL_n$ are also defined up to
residual gauge terms, and the reality of $\tG[T,S]$ is expressed by
demanding $L_{-n}=L_n^*$, $\bL_{-n}=\bL_n^*$.
Solving equation \eq{6.17} in terms of $L_n$, $\bL_n$, one finds:
\bea
&& 2L_n=-\tG[T=S=e^{inx^+}]  \nn\\
&& 2\bL_n=-\tG[T=-S=e^{inx^-}] \,.                        \lab{6.18}
\eea

Now, we can rewrite the canonical PB algebra \eq{6.12} in terms of the
new asymptotic generators. The definitions \eq {6.18} and the
relations \eq{6.9}, \eq{6.10} lead to
\bsubeq\lab{6.19}
\bea
&& \left\{L_n,\,L_m\right\}=
   -i(n-m)L_{n+m}-2\pi i\,a\ell\,n^3\d_{n,-m}\,,     \lab{6.19a}\\
&& \left\{\bL_n,\,\bL_m\right\}=
   -i(n-m)\bL_{n+m}-2\pi i\,a\ell\,n^3\d_{n,-m}\,.   \lab{6.19b}\\
&& \left\{L_n,\,\bL_m\right\}=0 \,,                  \lab{6.19c}
\eea
\esubeq
Upon the redefinition of the zero modes, $L_0\to L_0+\pi a\ell$,
$\bL_0\to \bL_0+\pi a\ell$, we obtain the standard form of the
Virasoro algebra with the {\it classical\/} central charge:
\bsubeq\lab{6.20}
\bea
&& \left\{L_n,\,L_m\right\}=
  -i(n-m)L_{n+m}-2\pi i\,a\ell\,n(n^2-1)\d_{n,-m}\,,\lab{6.20a}\\
&& \left\{\bL_n,\,\bL_m\right\}=
  -i(n-m)\bL_{n+m}-2\pi i\,a\ell\,n(n^2-1)\d_{n,-m}\,.\lab{6.20b}
\eea
\esubeq

Using the standard string theory normalization of the central charge,
we have
\be
c=12\cdot 2\pi a\ell=\frac{3\ell}{2G}\, .              \lab{6.21}
\ee
Thus, the value of the central charge in the theory of gravity with
torsion coincides with the Brown--Henneaux central charge of Einstein's
theory with cosmological constant, defined in Riemannian spacetime.

The form \eq{6.20} of the asymptotic algebra shows that
central term for the AdS subgroup, generated by the generators
$(L_{-1},L_0,L_1)$, $(\bL_{-1},\bL_0,\bL_1)$, vanishes. This is a
consequence of the fact that AdS subgroup is an exact symmetry of
the vacuum \eq{3.2}, \eq{3.3}, in agreement with the result of the
first reference in \cite{1}. The latter states that non-trivial
classical central term does not exist if at least one exact solution
is left invariant under the action of the asymptotic symmetry group.

\section{Concluding remarks}

We presented here an investigation of the structure of asymptotic
symmetry in 3d gravity with torsion. We have chosen a specific form of the
Baekler--Mielke action \cite{13,14} which yields the teleparallel dynamics,
in order to isolate the influence of torsion on the asymptotic dynamics.

Our procedure for constructing vacuum solutions is based on the
maximally symmetric form of the Riemannian piece of the curvature in
Riemann--Cartan geometry, equation \eq{2.5}. We obtained two exact
vacuum solutions, the AdS solution and the black hole, both in the
realm of the teleparallel geometry.

The results concerning the asymptotic symmetry of the theory are based
on a natural definition of asymptotically AdS behaviour of dynamical
variables. Canonical analysis of the related generators reveals the
necessity for an improvement of their form by the addition of
appropriate boundary terms, which are interpreted as the conserved
charges of the teleparallel theory. The canonical algebra of the
generators is realized as the Virasoro algebra with the classical
central charge $c=3\ell/2G$. The fact that the value of this charge is
the same as in Riemannian spacetime of general relativity indicates
that the boundary dynamics in 3d gravity depends much more on the form
of asymptotic conditions than on the underlying geometry.

\acknowledgements

This work was partially supported by the Serbian Science foundation,
Yugoslavia, and by the Slovenian Science foundation, Slovenia.

\appendix

\section{Symmetries of the AdS vacuum}

The invariance condition for the AdS vacuum \eq{3.2}, \eq{3.3} leads
to the following relations:
\bea
&&\xi^0{}_{,0}=-\frac{r}{\ell^2 f^2}\,\xi^1 \qquad
     \xi^1{}_{,1}=\frac{r}{\ell^2 f^2}\,\xi^1 \qquad
         \xi^2{}_{,2}=-\frac{1}{r}\,\xi^1              \nn\\
&&\xi^1{}_{,0}=f^4\xi^0{}_{,1} \qquad
     \xi^0{}_{,2}=\frac{r^2}{f^2}\,\xi^2{}_{,0} \qquad
         \xi^1{}_{,2}=-r^2 f^2\xi^2{}_{,1}            \nn\\
&&\th^0=-rf\xi^2{}_{,1} \qquad \th^1=\frac{f}{r}\,\xi^0{}_{,2}
   \qquad \th^2=-\frac{1}{f^2}\,\xi^1{}_{,0}           \nn
\eea
The general solution of these equations has the form
\bea
&&\xi^0=\ell \s_1-\frac{r}{f}\pd_2Q \,,\qquad
  \xi^1= \ell^2 f\pd_0\pd_2Q\,,\qquad
  \xi^2=\s_2-\frac{\ell^2 f}{r}\pd_0Q \,,                 \nn\\
&&\th^0=-\frac{\ell^2}{r}\,\pd_0Q\, ,\qquad
  \th^1=Q\, ,\qquad
  \th^2=\frac{1}{f}\,\pd_2Q\,,                            \lab{A1}
\eea
where
\be
Q\equiv \s_3\cos x^+ +\s_4\sin x^+
       +\s_5\cos x^- +\s_6\sin x^- \,,                    \lab{A2}
\ee
and $\s_i$ are six arbitrary dimensionless parameters.
For the basis of six independent Killing vectors we can take:
\bea
&&\xi_{(1)}=(\ell\,,\,0\,,\,0)\, ,\nn\\
&&\xi_{(2)}=(0\,,\,0\,,\,1)\, ,\nn\\
&&\xi_{(3)}=\left(\frac{r}{f}\,\sin x^+ \, ,\
            -\ell f\cos x^+ \, ,\
            \frac{\ell f}{r}\sin x^+ \right)  \nn\\
&&\xi_{(4)}=\left(\frac{r}{f}\,\cos x^+ \, ,\
            \ell f\sin x^+ \, ,\
            \frac{\ell f}{r}\cos x^+ \right)  \nn\\
&&\xi_{(5)}=\left(\frac{r}{f}\,\sin x^- \, ,\
            \ell f\cos x^- \, ,\
           -\frac{\ell f}{r}\sin x^- \right)  \nn\\
&&\xi_{(6)}=\left(\frac{r}{f}\,\cos x^- \, ,\
            \ell f\sin x^- \, ,\
            -\frac{\ell f}{r}\cos x^- \right)\,.         \lab{A3}
\eea

We have explicitly verified that the above expressions for $\xi^\m$ and
$\th^i$ fall into the class of asymptotic transformations \eq{4.5}. In
particular, six inequivalent solutions for the parameters $T$ and $S$
define the six Killing vectors \eq{A3} of the teleparallel AdS vacuum
solution.

\section{Asymptotic form of the constraints}

Using the adopted asymptotic conditions \eq{4.2}, the secondary
constraints
$$
R^i{}_{\a\b}\approx 0\,,\qquad
T^i{}_{\a\b}-\frac{2}{\ell}\,\ve^i{}_{jk}\,b^j{}_{\a}b^k{}_{\b}
\approx 0
$$
are rewritten in the form
\bsubeq\lab{B1}
\bea
&\om^0{_1}=\cO_4 \,,\qquad \om^2{_1}=\cO_4 \,,&         \lab{B1a}\\
&\big(r\a_2\big)_{,1}=\cO_3 \,, \qquad
  \big(r\b_2\big)_{,1} = \cO_3\,,&                      \lab{B1b}\\
&\ell\left(\Om^2{_2}-\Om^0{_2}\right)+r^2\Om^1{_1}
  =\left[r\left(B^2{_2}-B^0{_2}\right)\right]_{,1}
  +\cO_3 \,,&                                           \lab{B1c}\\
&B^0{_2}+B^2{_2}+\frac{r^2}{\ell}\,B^1{_1}
  -\ell\,\Om^0{_2}
  =\left(rB^2{}_2\right)_{,\,1}+ \cO_3\,,&              \lab{B1d}
\eea
\esubeq
where
\be
\a_{\m}\equiv \om^0{}_{\m}+\frac{1}{\ell}\,b^2{}_{\m}
-\frac{1}{\ell}\,b^0{}_{\m}\,,\qquad
\b_{\m}\equiv \om^2{}_{\m}+\frac{1}{\ell}\,b^0{}_{\m}
-\frac{1}{\ell}\,b^2{}_{\m}\,.                           \lab{B2}
\ee
From \eq{B1b}, we see that the terms ${\cal E}^{\a}$ and ${\cal
M}^{\a}$, included in the surface integrals, have the asymptotic
behaviour
$$
{\cal E}^1{}_{,\,1}=\cO_3 \,,\quad
{\cal E}^2=\cO_3 \,,\qquad\quad
{\cal M}^1{}_{,\,1}=\cO_3 \,,\quad
{\cal M}^2=\cO_3 \,,
$$
wherefrom one verifies the finiteness of the surface term $\G$ for
every region of integration (not necessarily a circle).

Beside the constraints, the equations of motion also refine the
asymptotic behaviour of the fields. This is because we demand that
the adopted asymptotics is conserved in time, which imposes
additional restrictions at infinity. Thus, one finds
\bsubeq\lab{B3}
\bea
&\ell\,\a_0+\b_2=\cO_3\,,\qquad
 \ell\,\b_0+\a_2=\cO_3\,,&                              \lab{B3a}\\
&\ell\,\Om^0{_0}+\Om^0{_2}=\cO_3\,,\qquad
 \ell\,\Om^2{_0}+\Om^2{_2}=\cO_3\,,&                    \lab{B3b}\\
&B^0{_0}+B^2{_0}+\frac{r^2}{\ell^2}\,B^1{_1}
  -\ell\,\Om^2{_0}
  =\left(rB^0{}_0\right)_{,\,1}+ \cO_3\,.&              \lab{B3c}
\eea
\esubeq
The terms $\Om^i{}_{\m}$ and $B^i{}_{\m}$ are defined in \eq{4.2}.

\section{Non-triviality of the boundary terms}

In this Appendix, we shall examine if some of the generators $\tG$
have trivial central terms, or equivalently, if the boundary term
$\G[T,S]$ vanishes on-shell for some values of the parameters $T$
and $S$.

First, note that all physically acceptable solutions of the equations
of motion are gauge equivalent to the black hole solution \eq{3.5},
\eq{3.6} \cite{21}. As our asymptotic conditions restrict the full
gauge group to the subgroup \eq{4.5}, we shall consider the set  of
solutions $\cal W$, defined by
$$
\begin{array}{ccl}
{\cal W}&=&\mbox{the black hole solution, plus all its transforms}\\
         &&\mbox{under the action of the gauge subgroup \eq{4.5}.}
\end{array}
$$
Thus, solving the equation $\G''=0$ in terms of ($T''$, $S''$) on the
space of solutions ${\cal W}$ is equivalent to solving the equation
\be
\bar\G''+\d_0'\bar\G''=0 \quad
\mbox{for all }(m,J)\mbox{ and }(T',S')\,.                \lab{C1}
\ee
The bar over $\G$ denotes that the boundary term is evaluated on the
two-parameter space of the black hole solutions \eq{3.5}, \eq{3.6}.

For $T'=S'=0$, we find the condition $\bar\G''=0$, or equivalently,
\be
\int T''d\vphi=\int S''d\vphi=0\,.                        \lab{C2}
\ee
Thus, the zero modes of the functions $T''$, $S''$ must
vanish. Next, using \eq{6.8}, equation \eq{C1} can be brought
to $\bar\G'''+C'''=0$ for all $(m,J)$ and $(T',S')$, which means that
$$
\int T'''d\vphi=\int S'''d\vphi=0=C''' \quad
\mbox{for all  } (T',S')\, .
$$
We now concentrate on the first two equations rewritten as
\[
\int d\vphi\left(S'T''_{,\,2}+T'S''_{,\,2}\right)=
\int d\vphi\left(S'S''_{,\,2}+T'T''_{,\,2}\right)=0\,.
\]
By adding and subtracting the two equations, they take the form
$$
\int d\vphi\,(T'+S')(T''+S'')_{,2}=
\int d\vphi\,(T'-S')(T''-S'')_{,2}=0
$$
for all allowed values of $T'$ and $S'$. Substituting here the general
solution \eq{4.7} for the parameters $T$ and $S$, one finds
$$
\int d\vphi\, f'f''_{,\,2}=
\int d\vphi\, g'g''_{,\,2}=0 \quad \forall\ f',g',
$$
which can hold only if $f''$ and $g''$ are constants. Consequently,
\be
S''=\mbox{ const},\quad T''=\mbox{ const}.               \lab{C3}
\ee
Taking into account the condition \eq{C2}, this result implies
\be
\G=0 \quad\Leftrightarrow\quad S=T=0 \,.                 \lab{C4}
\ee
Therefore, {\it only the trivial asymptotic generator $\tG[0,0]$ has
the vanishing surface term}. We can now be sure that the algebra
\eq{6.12} is valid {\it off-shell}, because there are no well defined
asymptotic generators that weakly vanish \cite{25}.


\begin{references}

\bibitem{1} J. D. Brown and M. Henneaux, Central Charges in the
  Canonical Realization of Asymptotic Symmetries: An Example from Three
  Dimensional Gravity, Comm. Math. Phys. {\bf 104} (1986) 207;
  see also M. Henneaux, Energy-momentum, angular momentum, and
  superscharge in 2+1 dimensions, Phys. Rev. {\bf D29} (1984) 2766.
\bibitem{2} E. Witten, 2+1 dimensional gravity as an exactly soluble
  system, Nucl. Phys. {\bf B311} (1988) 46;
  A. Achucarro and P. Townsend, A Chern-Simons Action For Three-Dimensional
  Anti-De Sitter Supergravity Theories, Phys. Lett. {\bf B180} (1986) 89.

\bibitem{3} O. Coussaert, M. Henneaux and P. van Driel, The asymptotic
  dynamics of three-dimensional Einstein gravity with negative cosmological
  constant, Class. Quant. Grav. {\bf 12} (1995) 2961.
\bibitem{4} M. Ba\~nados, Global charges in Chern-Simons theory and 2+1
  black hole, Phys. Rev. {\bf D52} (1996) 5861.
\bibitem{5} A. Strominger, Black hole entropy from Near--Horizon
  Microstates, JHEP {\bf 9802} (1998) 009.
\bibitem{6} J. Navaro--Salas and P. Navaro, A Note on Einstein Gravity
  on AdS$_3$ and Boundary Conformal Field Theory, Phys. Lett.
  {\bf B439} (1998) 262.
\bibitem{7} M. Ba\~nados, Three-dimensional quantum geometry and black
  holes, Invited talk at the Second Meeting ``Trends in Theoretical
  Physics", held in Buenos Aires, December, 1998 (hep-th/9901148).
\bibitem{8} M. Ba\~nados, Notes on black holes and three-dimensional
  gravity, Proceedings of the VIII Mexican School on Particles and Fields,
  AIP Conf. Proc. {\bf 490} (1999) 198.
\bibitem{9} M. Ba\~nados, T. Brotz and M. Ortiz,
  Boundary dynamics and the statistical mechanics of the 2+1 dimensional
  black hole, Nucl. Phys. {\bf B545} (1999) 340.
\bibitem{10} J. Zanelli, Chern-Simons Gravity: From 2+1 to 2n+1 Dimensions,
  Lectures presented at the XX Econtro de Fisica de Particulas e Campos,
  Brasil, October 1998, and at the Fifth La Hechicera School, Venezuela,
  November 1999, Braz. J. Phys. {\bf 30} (1999) 251.

\bibitem{11} F. W. Hehl, Four lectures on Poincar\'e gauge theory, in:
  Proceedings of the 6th Course of the School of Cosmology and Gravitation
  on {\it Spin, Torsion, Rotation and Supergravity\/,} held at Erice, Italy,
  1979, eds. P. G. Bergmann, V. de Sabbata (Plenum, New York, 1980) p. 5;
  E. W. Mielke, {\it Geometrodynemics of Gauge Fields\/} -- On the geometry
  of Yang--Mills and gravitational gauge theories (Academie--Verlag, Berlin,
  1987).
\bibitem{12} M. Blagojevi\'c, {\it Gravitation and gauge symmetries\/}
  (Bristol, Institute of Physics Publishing, 2001).

\bibitem{13} E. W. Mielke, P. Baekler,  Topological gauge model of
  gravity with torsion, Phys. Lett. {\bf A156} (1991) 399.
\bibitem{14} P. Baekler, E. W. Mielke, F. W. Hehl,
  Dynamical symmetries in topological 3D gravity with torsion,
  Nuovo Cim. {\bf B107} (1992) 91.

\bibitem{15} C. M\o ller, Mat. Fys. Scr. Dan. Vid. Selsk. {\bf 89}, No. 13
  (1978); K. Hayashi and T. Shirafuji, Phys. Rev. {\bf D19} (1979) 3524.
\bibitem{16} J. Nitsch, in {\it Cosmology and Gravitation: Spin,
  torsion, Rotation and Supergravity\/}, eds. P. G. Bergmann and V. de
  Sabbata (Plenum, New York, 1980) p. 63; F. W. Hehl, J. Nitsch and P. von
  der Heyde, in {\it General Relativity and Gravitation\/} -- One Hundred
  Years after the birth of Albert Einstein, ed. A. Held (Plenum, New York,
  1980) vol. 1, p. 329.
\bibitem{17} T. Kawai,  Teleparallel theory of (2+1)-dimensional gravity,
  Phys. Rev. {\bf D48} (1993) 5668;
  Poincar\'e gauge theory of (2+1)-dimensional gravity, Phys. Rev.
  {\bf D49} (1994) 2862;
  Exotic black hole solution in teleparallel theory of (2+1)-dimensional
  gravity,  Prog. Theor. Phys. {\bf 94} (1995) 1169-1174;
  A. A. Sousa, J. W. Maluf, Canonical Formulation of Gravitational
  Teleparallelism in 2+1 Dimensions in Schwinger's Time Gauge,
  Prog. Theor. Phys. {\bf 104} (2000) 531.

\bibitem{18} T. Regge and C. Teitelboim, Role of surface integrals in the
  Hamiltonian formulation of general relativity, Ann. Phys. (N.Y) {\bf 88}
  (1974) 286.

\bibitem{19} J. Zanelli, (Super-)Gravities Beyond 4 Dimensions, Lectures
given at the 2001 Summer School {\it Geometric and Topological Methods
for Quantum Field Theory\/}, Villa de Leyva, Colombia, June 2001,
e-print hep-th/0206169.

\bibitem{20} S. W. Hawking and G. F. R. Ellis, {\it The Large Scale
  Structure of Spacetime\/} (Cambridge, Cambridge Ubiversity Press, 1973).
\bibitem{21} M. Ba\~nados, C. Teitelboim and J. Zanelli, The Black Hole in
  Three-Dimensional Spacetime, Phys. Rev. Lett. {\bf 16} (1993) 1849;
  M. Ba\~nados, M. Henneaux, C. Teitelboim and J. Zanelli, Geometry of
  2+1 Black Hole, Phys. Rev. {\bf D48} (1993) 1506.
\bibitem{22} A. Garc\'\i a, F. W. Hehl, C. Heinecke and A. Mac\'\i as,
  Exact vacuum solutions of (1+2)-dimensional Poincar\'e gauge theory: BTZ
  solution with torsion, Cologne preprint (in preparation), 2002.
\bibitem{23} M. Henneaux and C. Teitelboim, Asymptotically Anti-de Sitter
  Spaces, Commun. Math. Phys. {\bf 98} (1985) 391.
\bibitem{24} L. Castellani, Symmetries of constrained Hamiltonian systems,
  Ann. Phys. (N.Y) {\bf 143} (1982) 357.
\bibitem{25} J. D. Brown and M. Henneaux, On the Poisson bracket of
  differentiable generators in classical field theory, J. Math. Phys.
  {\bf 27} (1986) 489.

\end{references}
\end{document}